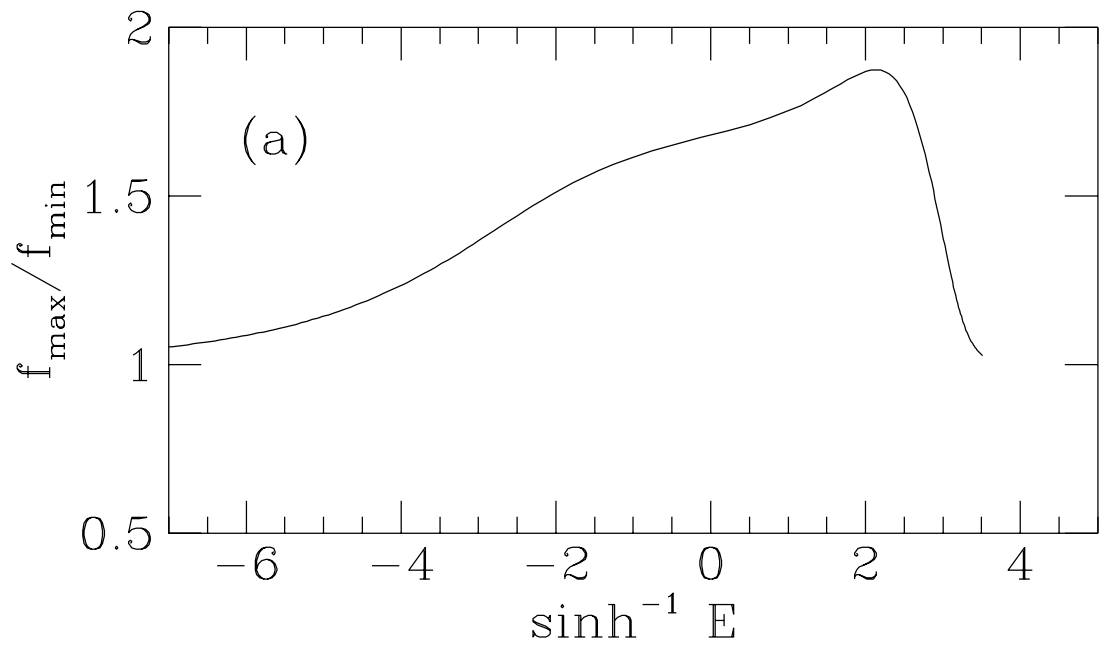

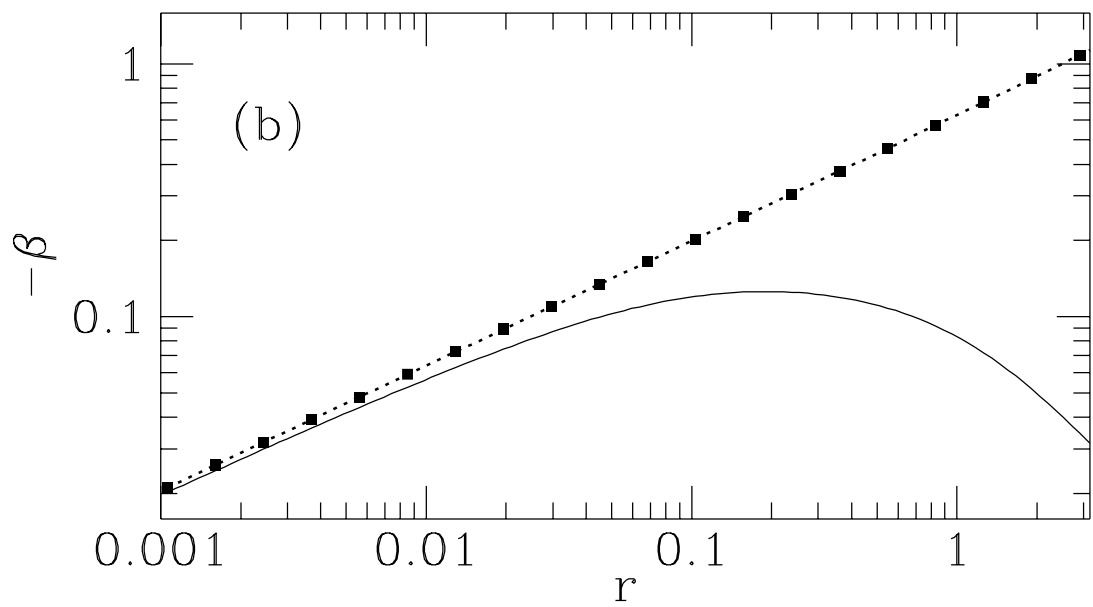

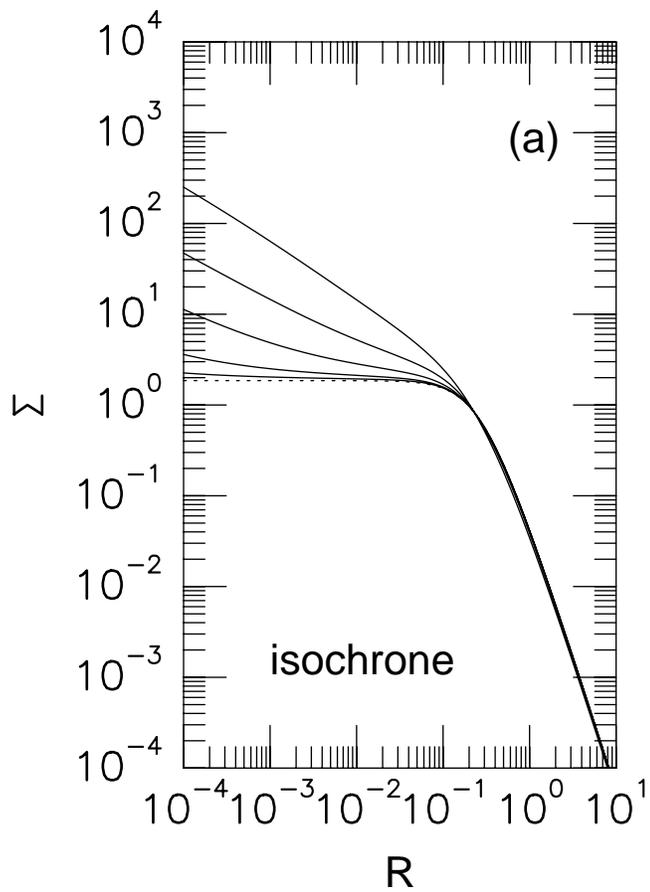
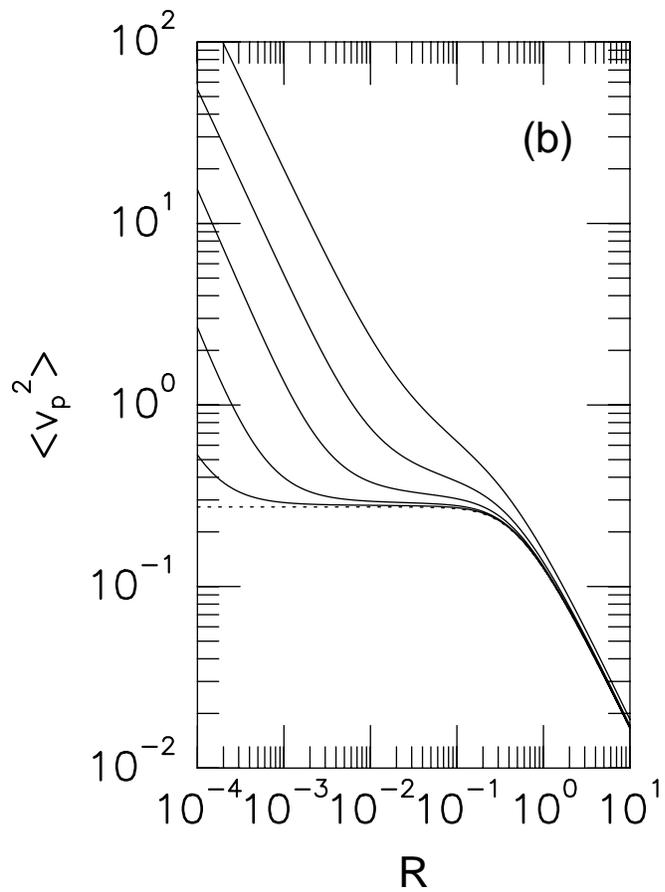
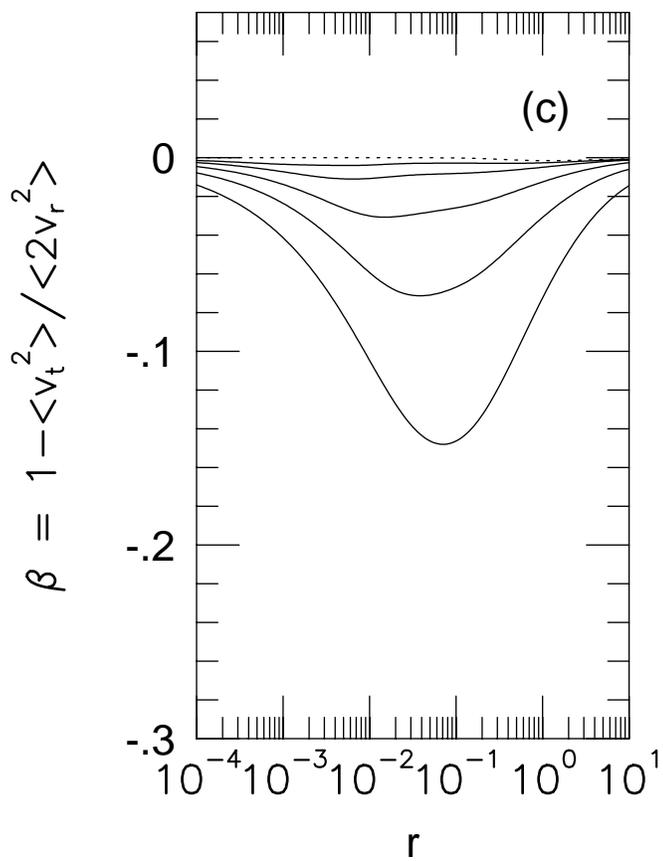
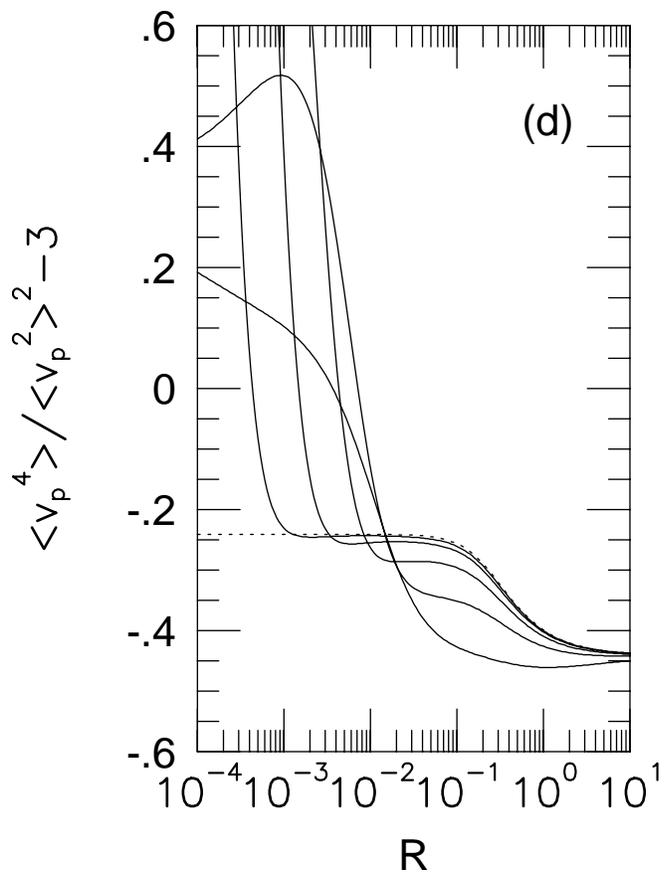

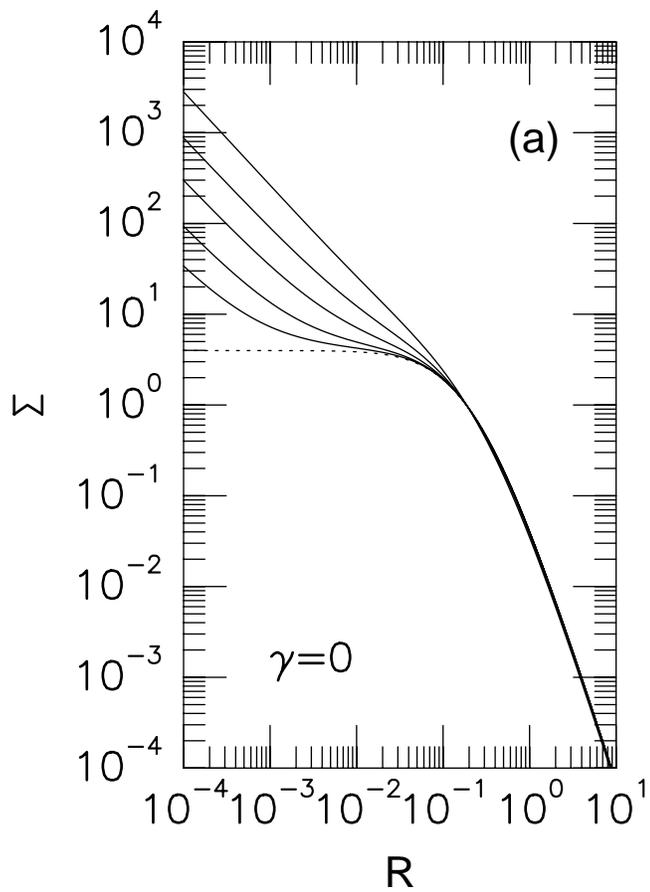
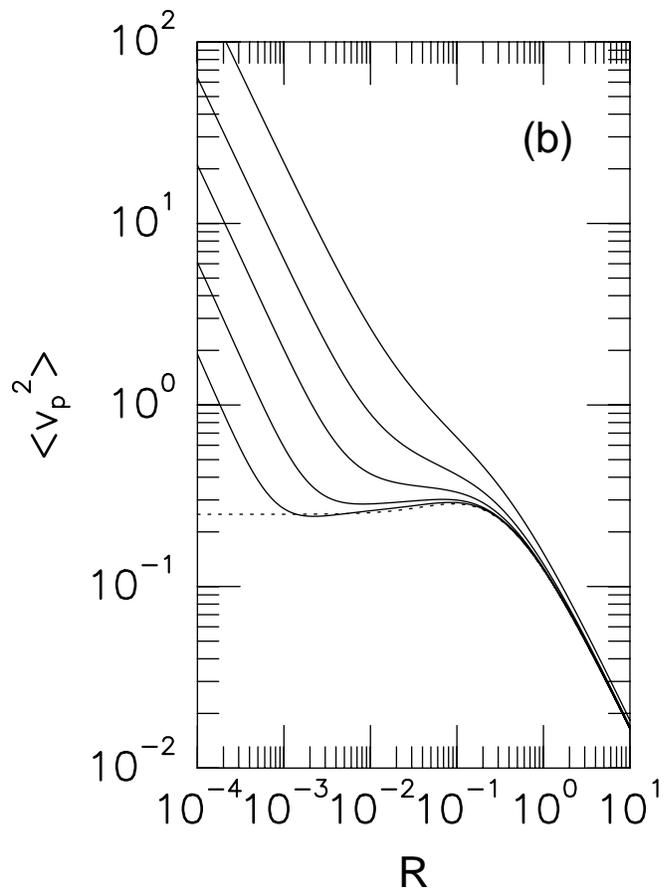
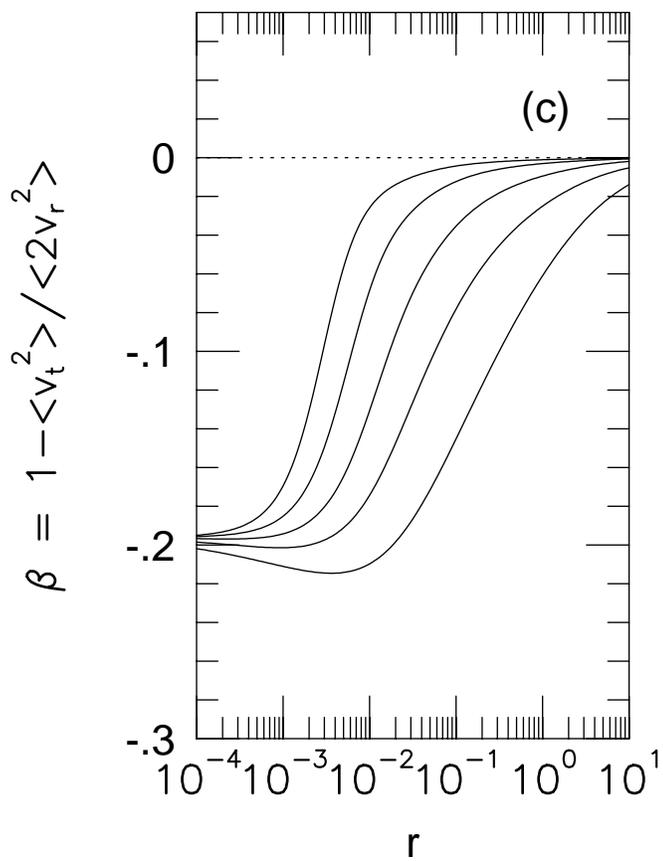
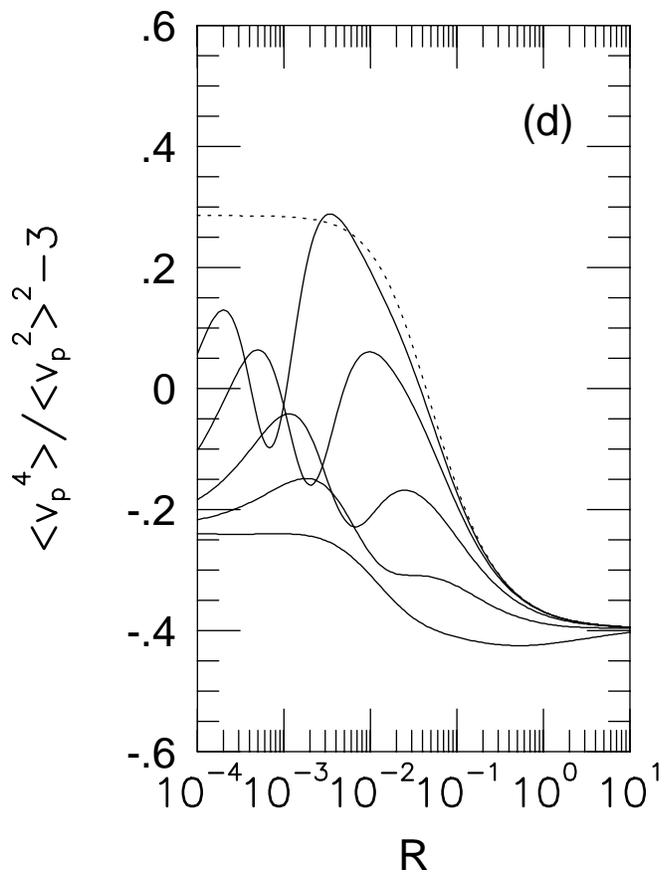

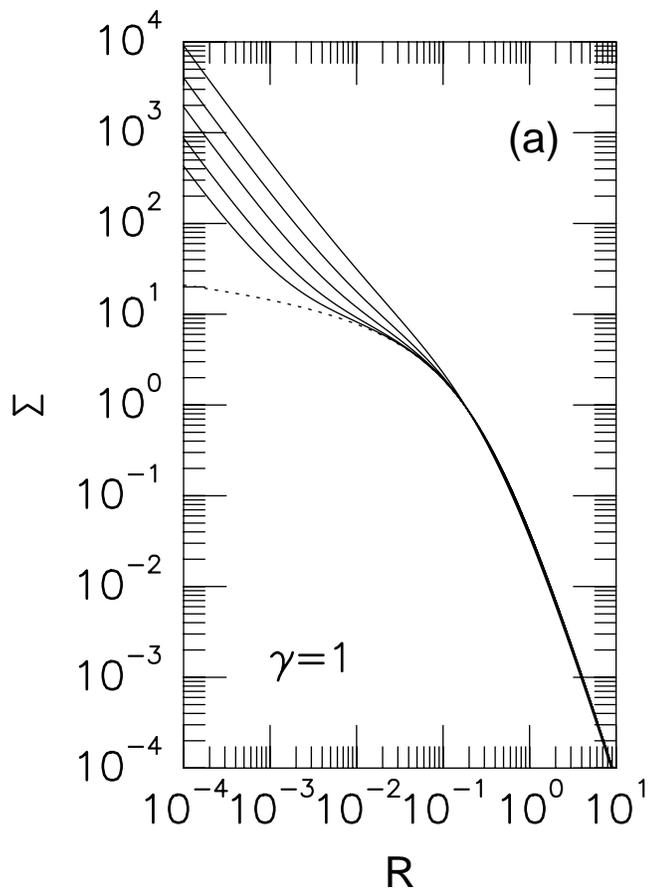
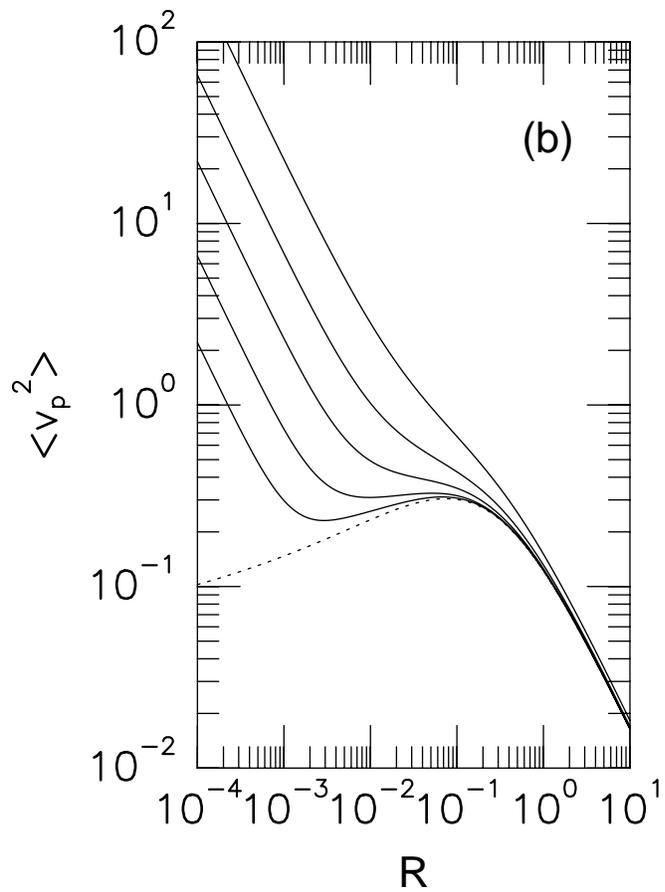
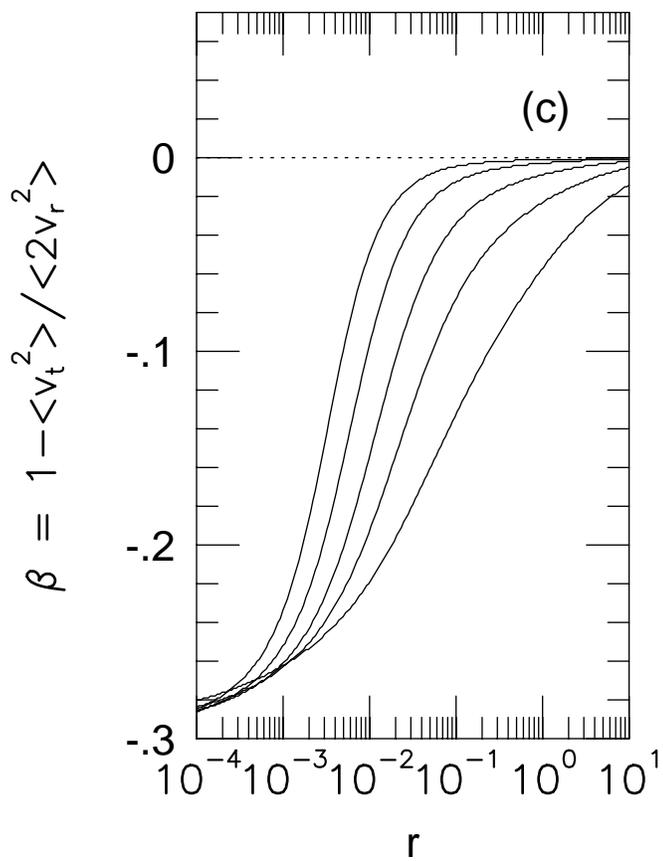
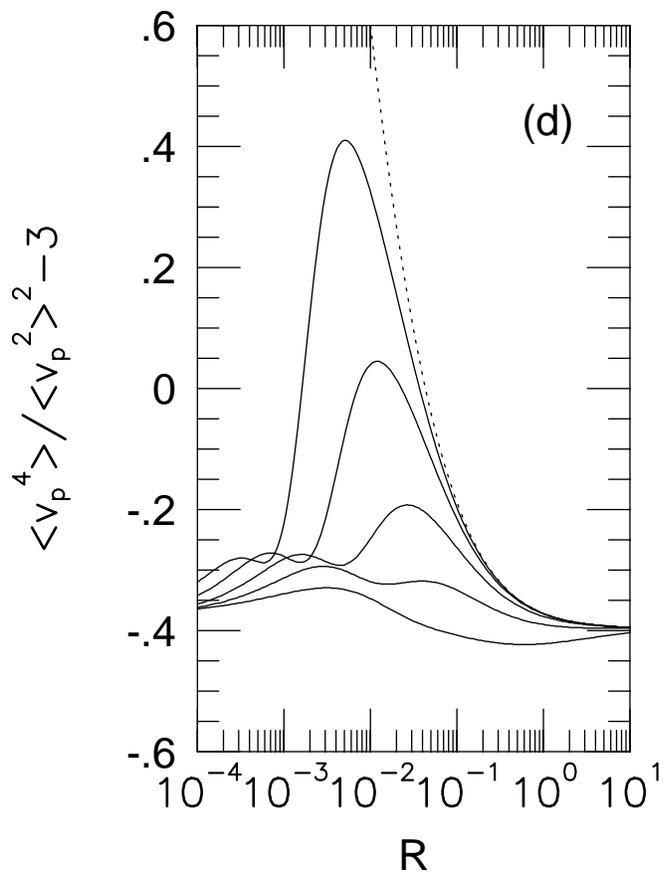

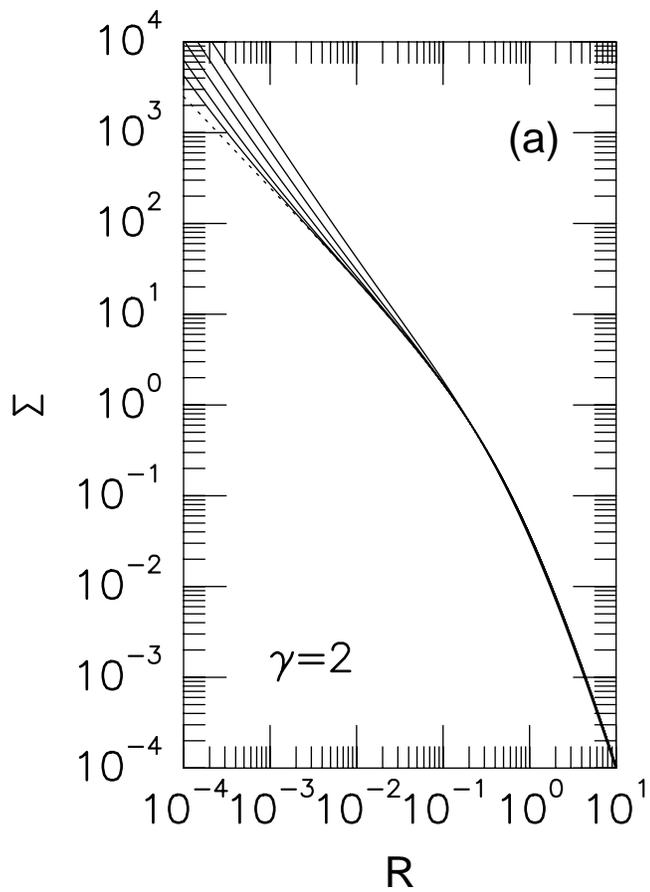
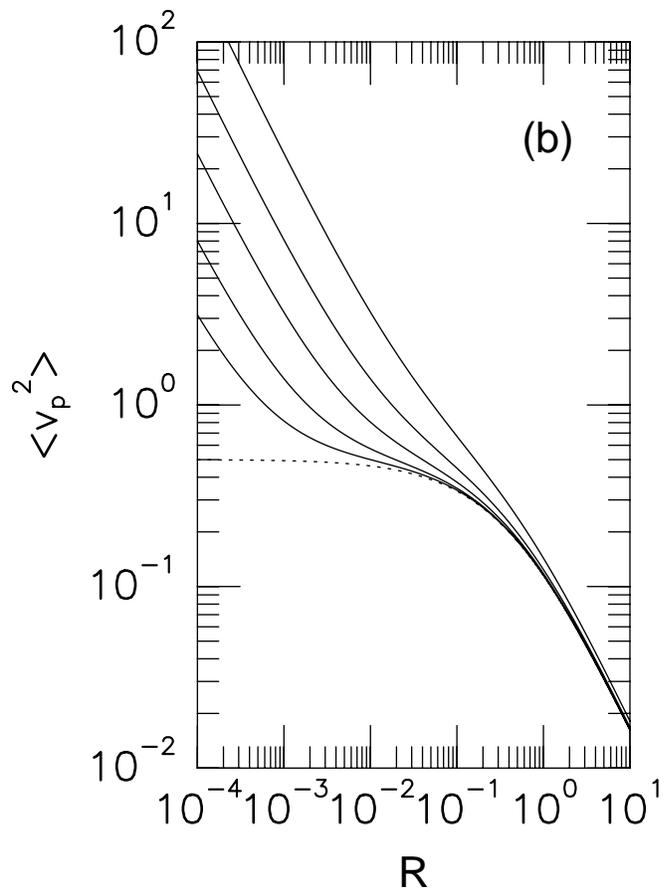
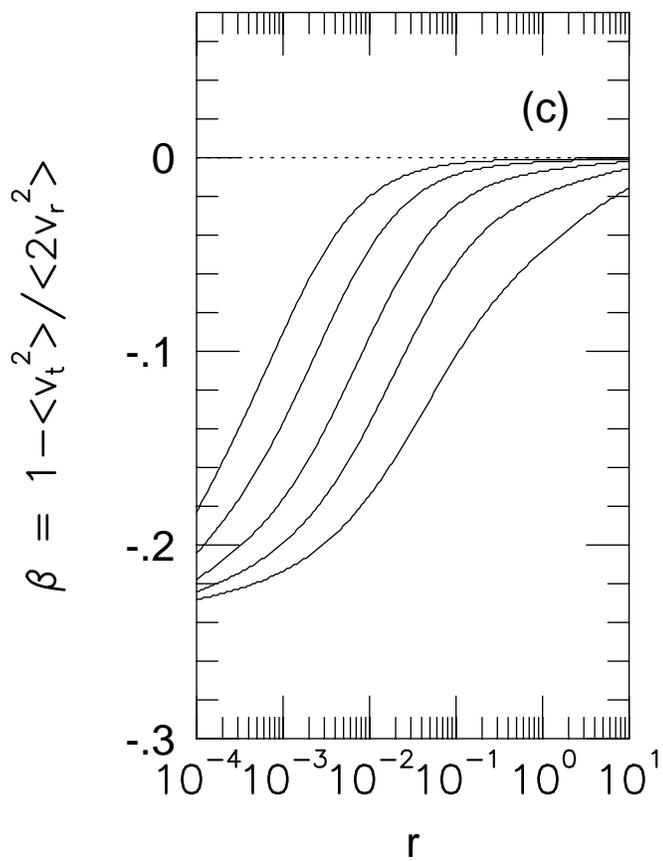
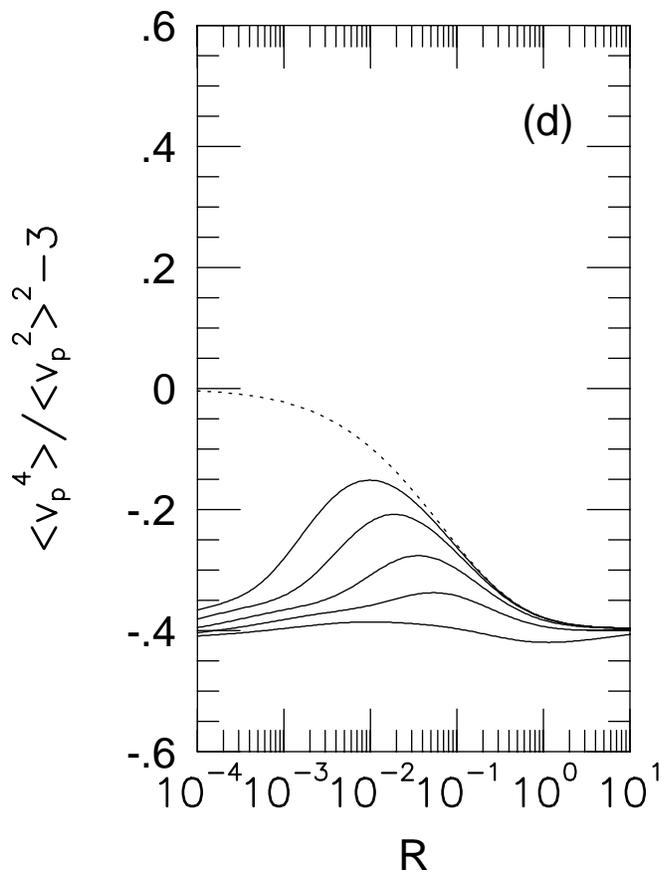

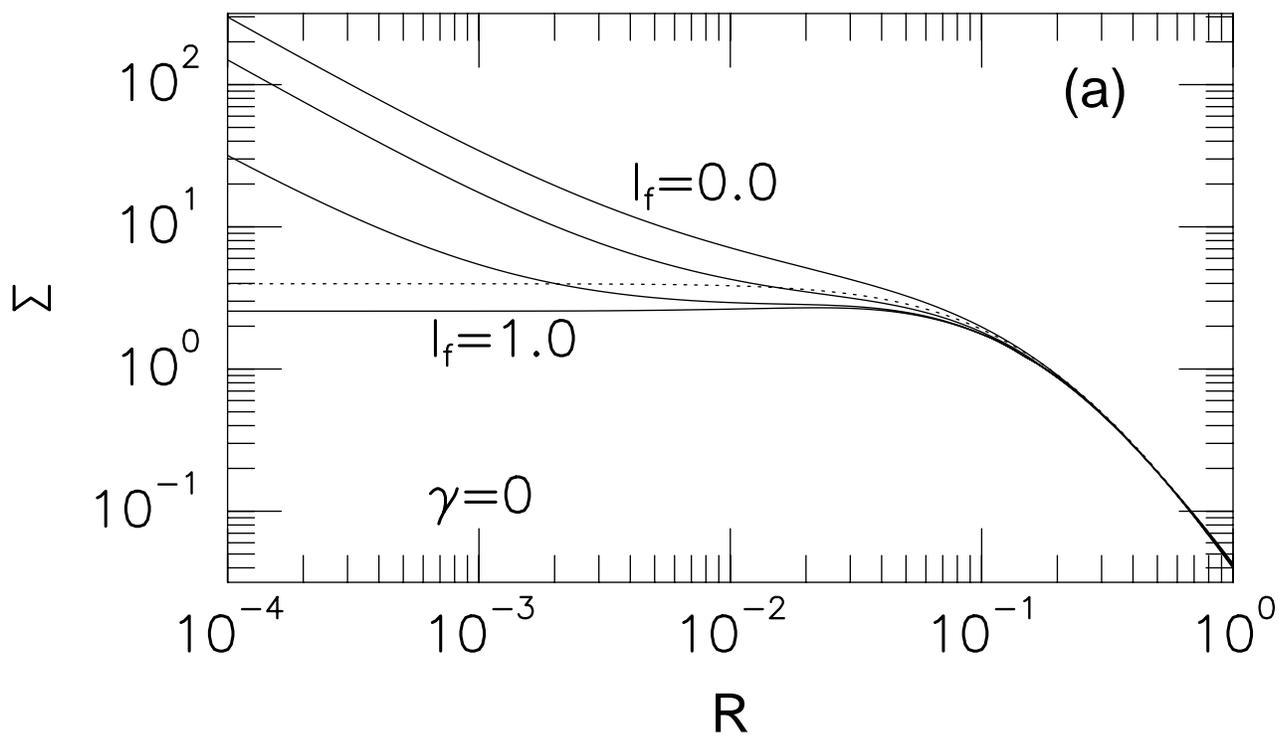

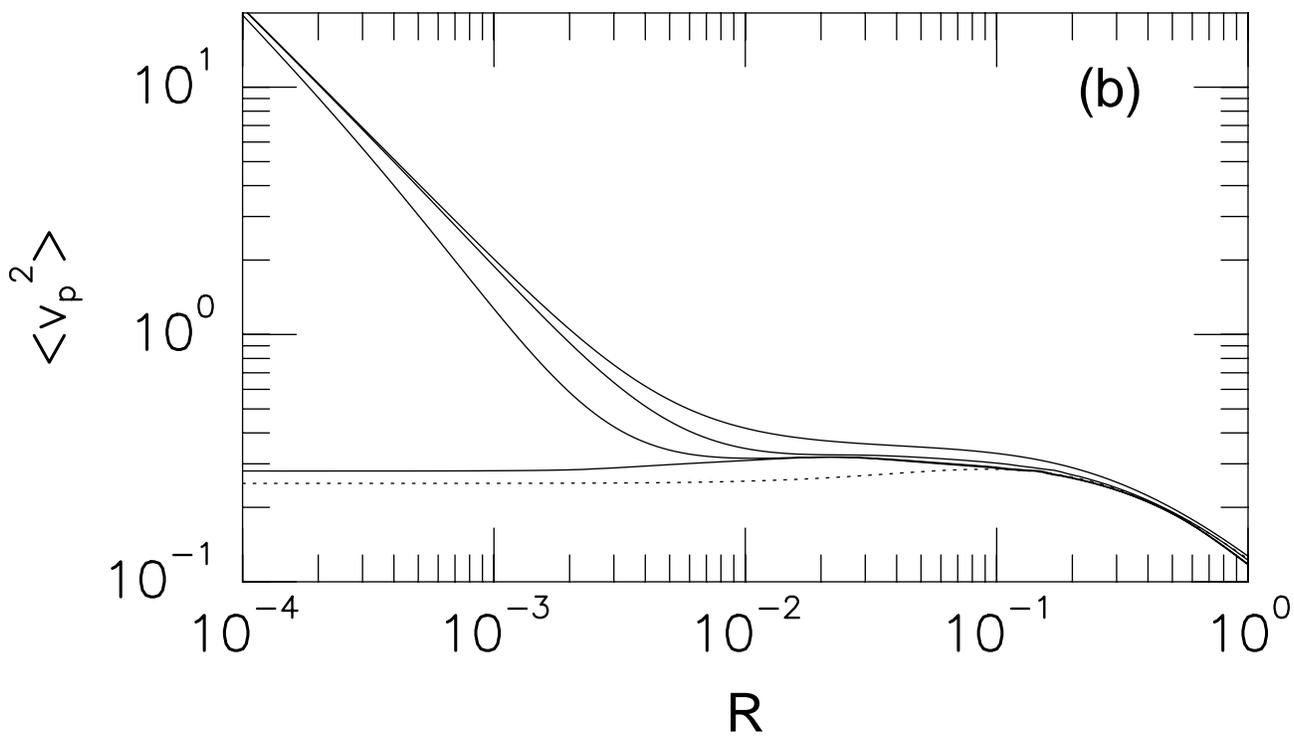

# Models of galaxies with central black holes: adiabatic growth in spherical galaxies

Gerald D. Quinlan, Lars Hernquist, and Steinn Sigurdsson

*Board of Studies in Astronomy and Astrophysics*
*University of California, Santa Cruz, CA 95064*

## ABSTRACT

We add black holes to nonrotating, spherical galaxy models, with the assumption that the black-hole growth is slow compared with the dynamical time but fast compared with the relaxation time. The outcome differs depending on whether the core of the initial galaxy does or does not resemble that of an isothermal sphere. For the isothermal case the previously-known results are confirmed and sharpened: the black hole induces cusps in the density ($\rho \sim r^{-3/2}$) and velocity dispersion ($v^2 \sim r^{-1}$), and a tangential anisotropy in the velocity distribution away from the center. For the non-isothermal case the induced density cusp is steeper, and the induced anisotropy is larger and penetrates right to the center. The cusp around the black hole is insensitive to anisotropy in the initial velocity distribution, and also to the origin of the black hole, unless its mass comes exclusively from the stars of lowest angular momentum, in which case the cusp is suppressed. We discuss the implications for the interpretation of evidence for massive black holes in galactic nuclei.

## 1. Introduction

The growth of central mass concentrations appears to be a natural result of the processes that shape large stellar systems. Quasars and active galactic nuclei are believed to derive their power from the most spectacular outcome of this: matter at the center of a galaxy that has collapsed into a massive black hole (BH). If this view is correct then many galaxies should today contain "dead quasars," massive BHs starved of fuel (Rees 1990). The search for dynamical evidence of BHs from the distribution and kinematics of stars in the centers of galaxies has thus received considerable attention, as constraints on BH masses would help us understand the structure of quasars and, more generally, the formation and evolution of dense galactic nuclei. There is evidence from ground-based observations that BHs have been detected in this way in a half-dozen or so nearby galaxies (see the reviews by Dressler 1989, Gerhard 1992, and Kormendy 1992), although a watertight case has yet to be made. Some of the remaining questions might be answered by high-resolution observations with the refurbished Hubble Space Telescope (HST); others could perhaps be answered by better theoretical modelling of existing data. In this paper we avoid detailed modelling of particular galaxies, and focus instead on a complementary approach — the construction of general theoretical models of galaxies with central BHs — in the hope that this might help both in identifying and measuring BHs in galaxies, and in explaining the origin of these systems.

By the construction of models of galaxies with central BHs we mean the exploration of equilibrium solutions that result from plausible initial conditions and definite (though perhaps speculative) BH formation scenarios. We thus exclude models in which a BH is placed at the center of a stellar system without regard to the origin of that configuration, with simplifying assumptions made without justification. Examples of these include the "loaded polytropes" of Huntley and Saslaw (1975), which assume an isotropic velocity distribution and a polytropic relation between the pressure and density, and the "$\eta$-models with BHs" of Tremaine et al. (1994), which assume an isotropic velocity distribution and a law for the variation of density with radius. These models can give helpful mathematical insights into the range of allowed solutions, but their arbitrary nature makes it unlikely that they will match real galaxies well. We also exclude models constructed by techniques such as linear programming to match observations of galaxies believed to contain BHs (see, e.g., Richstone and Tremaine 1985). These models are valuable for deciding whether particular galaxies contain BHs (and for constraining the BH masses if they do), but they are not based on initial conditions and a BH formation scenario, and hence can neither help us assess theories for these nor make general predictions for the solutions we should expect to find.

The expected distribution of stars around a massive BH was studied in detail in the late 1970s and early 1980s, but in the context of globular clusters, not galactic nuclei (see Shapiro 1985 for a review). The key assumption underlying this work is that the two-body relaxation time is short compared with the age of the system. The cluster is then driven by relaxation processes to a steady-state solution in which the consumption of stars by the BH and the diffusion of new stars into the loss cone result in a density cusp $\rho(r) \sim r^{-7/4}$ (Bahcall and Wolf 1976).

In most galaxy cores, however, the relaxation time is long compared with the age of the system, and there is no reason to expect a unique solution. The distribution of stars around a central BH in a galaxy is likely to depend upon many factors, including the order in which the galaxy and the BH form, the structure of the galaxy — assuming that it formed first — before the BH forms (spherical, axisymmetric, or triaxial? rotating or nonrotating? cuspy or flat? isotropic or anisotropic?), and the origin of the BH mass (stars? gas? an external source?). If we had predictions for the cusps that result from all the possibilities we would have a good understanding of what we might find, and could rule some possibilities out by comparing their predictions with observations.

Only one plausible formation scenario has been explored in detail: the growth of a central BH from the accumulation of gas on a time scale long enough that the stellar action variables are adiabatically conserved. The consequences of this can be derived most easily for spherical galaxies. Peebles (1972) considered the adiabatic growth of a central BH in an isothermal sphere, and showed it would lead to a density cusp $\rho(r) \sim r^{-3/2}$. Young (1980) constructed numerical models that confirmed Peebles's result and showed that the BH induces a tangential anisotropy in the velocity distribution. Goodman and Binney (1984) obtained an approximate solution to the same problem, and showed that the velocity distribution remains isotropic at the center despite the tangential bias induced nearby. Lee and Goodman (1989) generalized Young's calculation in an approximate way to study the growth of BHs in axisymmetric, rotating galaxies.

The adiabatic growth of a BH in a triaxial galaxy is more difficult to analyse than the corresponding problem for a spherical galaxy, but is potentially more interesting because the maintenance of triaxiality depends on the existence and selective population of box orbits, orbits that — given enough time — pass arbitrarily close to the central BH. Most studies of this problem have considered only the effect of a BH on single-particle orbits, and have not attempted to follow the self-consistent evolution of a whole galaxy. Gerhard and Binney (1985) argued that the scattering and redistribution of box orbits by a BH would force the inner regions of a triaxial galaxy to become rounder, out to a distance of about 1 kpc for a $10^8 M_\odot$ BH in a giant elliptical galaxy, perhaps resulting in a global change in the galaxy shape. Similar arguments were made by Hasan and Norman (1990) for the growth of BHs in barred galaxies. Norman, May, and van Albada (1985) pioneered the use of N-body simulations to follow the self-consistent evolution of triaxial galaxies with growing BHs, and found results consistent with the predictions of Gerhard and Binney. The number of particles in their simulations was small (5000-20000), however, which limited the realism with which they could model the systems, and which caused spurious relaxation that influenced the results in an uncertain way.

Despite these advances, many questions remain about the effect a growing central BH has on the structure of a galaxy. This is true even for spherical galaxies. We don't know what extremes are possible, e.g., how steep or gradual the density and velocity cusps can be, and whether it is possible to hide a BH at the center of a galaxy without an observable cusp. Nobody has made accurate predictions for the tangential anisotropy expected near a BH, which we need to interpret observations and to make precise mass estimates. We don't know exactly what signature a BH will leave in the distribution of velocities along the line-of-sight, a pressing question now that an effort is underway to quantify the deviation of these distributions from Gaussians and to use that information to constrain dynamical models (van der Marel et al. 1994). Another question is to what extent a BH can suppress instabilities such as the radial-orbit instability that plague some models for galactic nuclei without BHs (see, e.g., Merritt 1987; Palmer and Papaloizou 1988).

For non-spherical galaxies the questions are more numerous and profound. The main question is whether a central BH is compatible with triaxiality, i.e., whether a the growth of a BH in a triaxial galaxy forces the galaxy to become rounder and, if it does, whether this happens gradually from the inside out, or abruptly in a manner that affects the whole galaxy. The answer might allow us to constrain BH masses and formation histories from observed isophote shapes, and might help explain why unresolved elliptical galaxy cores tend to appear disky while resolved ones tend to appear boxy (Nieto, Bender, and Surma 1991). Figure rotation and resonances can mitigate the influence of the BH and complicate the analysis by preventing orbits from approaching close to the center (Pfenniger and de Zeeuw 1989). The growth of a central BH in a triaxial galaxy will cause many orbits that were regular to become stochastic, although it is not clear what stochasticity implies for the structure of the galaxy (Udry and Pfenniger 1988). Many of these questions are difficult and can be addressed only with the help of large N-body experiments.

In this paper we start with the simplest problem, and study the adiabatic growth of a central BH in a nonrotating, spherical galaxy using the numerical approach of Young (1980). There are several reasons for re-examining this problem. The first is that previous work considered the growth of a BH in only one galaxy model, the isothermal sphere, and it is not clear which of the results are peculiar to this model and which are general. We now know that few elliptical-galaxy cores resemble that of the isothermal sphere; many have surface brightnesses that continue to rise at the smallest observable radii. We examine a family of simple galaxy models with this property, and show that the adiabatic growth of a central BH gives results for them that differ qualitatively from those for the isothermal sphere. In particular, the density cusp induced by the BH is steeper, and the anisotropy in the velocity distribution is larger and penetrates right to the center.

Another reason for re-examining this problem is that previous work did not consider the origin of the BH, and assumed either that its mass came from a source external to the stars (such as gas that seeps in from the outer parts of the galaxy), or, if the BH did grow at the expense of the stars, that the mass loss could be ignored. We perform calculations both with and without taking stellar mass loss into account, and show that the mass loss is indeed unimportant unless it is highly concentrated towards the galaxy center.

Finally, we are re-examining this problem to extract quantitative results that were ignored or considered only qualitatively in previous work, such as the the anisotropy in the velocity dispersion, and the fourth moment of the line-of-sight velocity distribution. We are doing this partly because of their importance for the interpretation of observations, but also because we want to use them to calibrate an N-body program we are developing to study the growth of BHs in triaxial galaxies, a problem that cannot be handled by the simple techniques used here.

We assume that the mass concentration at the galaxy center is a massive BH, although some of our conclusions will apply to galaxies with other mass concentrations, such as star clusters, provided that they are sufficiently dense.

## 2. Computational methods

### 2.1. Strategy

The computational strategy and equations on which our calculations are based are described in detail by Young (1980), and are summarized only briefly here.

The starting point of the calculation is an equilibrium spherical galaxy with no BH (or a small BH). Young then adds a BH to the center making two assumptions: first, that the BH mass comes from a source external to the stellar system, and does not deplete the stellar distribution function; second, that the BH growth is slow enough that the stellar action variables (radial action $J_r$ and angular momentum $L$) are adiabatically conserved, but fast enough that two-body relaxation can be ignored. The first assumption is not essential and can be relaxed; we do this later in the paper and find that the results are not highly sensitive to the origin of the BH mass. The second assumption is necessary in this type of calculation, but is justified provided that the relaxation time is long compared with the age of the system, and that the BH growth is slow compared with the orbital period near the galaxy center.

Young's approach does not follow the time-evolution of the system, but solves directly for the final self-consistent distribution of stars in which the distribution function has the same dependence on the action variables as it did in the original model. This is done by adding a central BH and then passing repeatedly through the following loop to converge onto the solution: compute the potential from the mass distribution (including the BH); compute the action variables in that potential; adjust the distribution function $f(E,L)$ to remain a fixed function of the actions; compute the density generated by the distribution function in the current potential; check how much the density has changed from the previous iteration; decide whether to accept the solution or to pass through the loop again. A dozen or so repetitions are usually enough to reduce the change in density to less than one part in $10^4$ at all radii, which is our convergence criterion.

### 2.2. Computer program and output

Our program that implements this strategy is simple and easy to run, yet flexible enough to handle a variety of galaxy models. The density, distribution function, and other properties of the galaxy are described by their values on a discrete set of grid points: $\rho_i = \rho(r_i)$, $f_{ij} = f(E_i, L_j)$, etc. The radial grid points are spaced logarithmically between minimum and maximum values chosen by the user (typically $r_{\min} = 10^{-4}$ and $r_{\max} = 10^2$ in our units). The grid points for energy are chosen to match the potential at the radial grid points, and thus vary during the iterative procedure to converge onto the new potential. The grid points for angular-momentum are spaced linearly between $x_{\min} = 0$ and $x_{\max} = 1$, where $x = L/L_c$ is the ratio of the angular momentum to the circular angular momentum at the given energy. For the calculations in this paper we used 200 grid points for radius and energy and 20 for angular-momentum, although that is more than is necessary: the gross properties of the stellar cusps can be reproduced easily with half this number. A typical calculation takes about one minute to complete on our IBM 580 RISC workstation.

At the end of the calculation we have the complete distribution function for the stellar system, which we condense into a small number of moments as a function of radius ($r$) or projected radius ($R$). We first compute the density

$$\rho(r) = 4\pi \int_{\phi(r)}^{\phi(\infty)} dE \int_0^{L_m} dL\, L f(E,L)/(r^2 v_r), \quad (1)$$

and some low-order velocity moments ($m$ and $n$ are

even integers)

$$\langle v_r^m v_t^n \rangle(r) = \frac{4\pi}{\rho(r)} \int_{\phi(r)}^{\phi(\infty)} dE \int_0^{L_m} dL\, L f(E,L) \frac{v_r^m v_t^n}{r^2 v_r}, \quad (2)$$

where $L_m$ is the maximum angular momentum attainable by an orbit of energy $E$ at radius $r$,

$$L_m = \left[2r^2(E - \phi(r))\right]^{1/2}, \quad (3)$$

and $v_r$ and $v_t$ are the radial and tangential velocities,

$$v_r = \left[2(E - \phi(r)) - L^2/r^2\right]^{1/2}, \qquad v_t = L/r. \quad (4)$$

We then project the intrinsic density and velocity moments onto the plane of the sky to get the surface density

$$\Sigma(R) = 2 \int_R^\infty \frac{dr\, r\rho}{\sqrt{r^2 - R^2}}, \quad (5)$$

the dispersion of the line-of-sight velocity distribution (LOSVD)

$$\langle v_p^2 \rangle(R) = \frac{2}{\Sigma(R)} \int_R^\infty \frac{dr\, r\rho}{\sqrt{r^2-R^2}} \left[\left(1 - \frac{R^2}{r^2}\right)\langle v_r^2\rangle + \frac{1}{2}\frac{R^2}{r^2}\langle v_t^2\rangle\right], \quad (6)$$

and the fourth moment of the LOSVD (Merrifield and Kent 1990)

$$\langle v_p^4 \rangle(R) = \frac{2}{\Sigma(R)} \int_R^\infty \frac{dr\, r\rho}{\sqrt{r^2-R^2}} \left[\left(1 - \frac{R^2}{r^2}\right)^2\langle v_r^4\rangle + 3\frac{R^2}{r^4}(r^2 - R^2)\langle v_r^2 v_t^2\rangle + \frac{3}{8}\frac{R^4}{r^4}\langle v_t^4\rangle\right], \quad (7)$$

which we usually present as the dimensionless kurtosis $\kappa = \langle v_p^4\rangle/\langle v_p^2\rangle^2$. We also compute the anisotropy parameter

$$\beta = 1 - \langle v_t^2\rangle/\langle 2v_r^2\rangle, \quad (8)$$

which is 0 for an isotropic distribution and can vary between $-\infty$ (for purely circular orbits) and $+1$ (for purely radial orbits).

Some of our output quantities can be compared directly with observations and some cannot. The anisotropy parameter $\beta$ cannot, but it is nevertheless important because it must be known (or assumed) before observed values for $\Sigma$ and $\langle v_p^2\rangle$ can be converted into a precise mass estimate. The velocity moments $\langle v_p^2\rangle$ and $\langle v_p^4\rangle$ could be compared directly with observations if we could observe without noise and with infinite resolution, but in practice we cannot and the moments are affected by noise in the wings of the distribution (this is especially true for $\langle v_p^4\rangle$, which cannot be measured reliably) and by the nonzero seeing radius over which the observations are averaged. Gerhard (1993) and van der Marel and Franx (1993) argue that it is better to quantify observed LOSVDs by a set of Gauss-Hermite moments, which describe the LOSVD by a Gaussian fit and the deviations from this fit, and which are not as sensitive to the wings of the distribution as are moments such as $\langle v_p^2\rangle$ and $\langle v_p^4\rangle$. It might have been interesting to present our output in this form, but that would have made the program more complicated. The classical moments $\langle v_p^2\rangle$ and $\langle v_p^4\rangle$ are easy for us to compute and are sufficient to give an understanding of the intrinsic dynamics. For distributions close to a Gaussian, the Gauss-Hermite moment $h_4$ of van der Marel and Franx (1993) is related to the kurtosis by

$$\kappa \simeq 3 + 8\sqrt{6} h_4 \quad (9)$$

(this linear relation is unreliable if $|h_4| \gtrsim 0.03$).

### 2.3. Tests of program

To test the program we first ran it with simple galaxy models (described in Section 3.1.) with no BH to check that the output quantities matched those of the input models to sufficient accuracy (at least one part in $10^4$, except at the outermost radial grid points where the accuracy is worse) and that the accuracy improved in the expected manner when the number of grid points was doubled or quadrupled. To check the calculations of the anisotropy parameter $\beta$ and kurtosis $\kappa$ we used the anisotropic Plummer models of Dejonghe (1987) and Cuddeford (1991), and several other anisotropic models that we derived by Cuddeford's technique.

The main test of the program was to reproduce Young's (1980) results for the adiabatic growth of a BH at the center of an isothermal sphere. A visual comparison of our results with his showed satisfactory agreement. The only quantity for which we could detect any disagreement is the ratio $f_{\max}/f_{\min} = f(E, L_c(E))/f(E, 0)$, which Young plotted in his Figure 1 for just one BH mass (the largest he considered) to quantify the anisotropy in the distribution function. We show our version of this plot in Figure 1(a), in the same units used by Young. Our $f_{\max}/f_{\min}$ ratio approaches unity towards the left of the plot slightly slower than Young's does, but the difference is small and we do not view it as significant (our result does not change if we double or quadruple the grid resolution).

It is unfortunate that Young presented the anisotropy in this way, as it is not obvious how large an anisotropy it implies for the velocity dispersion (although Young concluded correctly that even for his largest BH mass the anisotropy was small). We show in Figure 1(b) the anisotropy parameter $\beta$ for the same calculation as in Figure 1(a). Note first that the anisotropy is small, as Young concluded, and second that it goes to zero at the center, as predicted by Goodman and Binney (1984). The Goodman-Binney solution[1] gives an anisotropy that varies with radius as $\beta \sim r^{1/2}$, which we show in Figure 1(b). This solution is expected to be valid only near the galaxy center, as it is derived by approximating the true potential by a harmonic potential before the BH is added and by a Kepler potential after. Binney and Petit (1989) estimate the boundary to the solution's validity to be the radius at which the enclosed stellar mass in the initial model equals the mass of the BH, but for the calculation shown in Figure 1 we find that at this radius ($r \simeq 0.6$) the anisotropy implied by the solution is about five times too large.

The dotted line in Figure 1(b) shows our program output when we turn off the self-consistent potential calculation and impose harmonic and Kepler potentials at all radii (chosen to match the central potentials before and after the BH is added). The output agrees with the Goodman-Binney solution, confirming that our program is correctly keeping the distribution function a fixed function of the action variables.

## 3. Results

### 3.1. Initial models and output figures

As the starting point for our calculations we pick well-known, simple galaxy models with mass distributions that span the range of behaviours expected for spherical galaxies. The models are not intended to be accurate representations of real galaxies, although two of them (the $\gamma = 1$ and $\gamma = 2$ models) do give reasonable fits to an $R^{1/4}$ law.

The first model we pick to have a core like that of the isothermal sphere. We do not use the isothermal sphere, because it was studied in detail by Young (1980). Instead we use Hénon's (1960) isochrone model, defined by the potential

$$\phi_I(r) = -\frac{GM}{b + \sqrt{b^2 + r^2}}. \quad (10)$$

The density corresponding to this potential falls off at large radii as $r^{-4}$. At small radii the density is nearly constant, and can be expanded in even powers of $r$:

$$\rho(r) \approx \rho(0) + \frac{1}{2}\rho''(0)r^2 + \ldots . \quad (11)$$

Models that share this property — such as the Plummer model, King models, and the isothermal sphere — are often called models with isothermal cores, a name we dislike because the word isothermal should refer to the velocity distribution, not the density, and because the singular isothermal sphere has a steep density cusp ($\rho \sim r^{-2}$) and yet certainly deserves to be called isothermal. It is nevertheless useful to have a name for galaxy models with the property (11), because, as we shall soon see, they all respond in a similar manner to the adiabatic growth of a central BH. We shall call them models with *analytic* cores[2], because a density with spherical symmetry must be expandable about the center as in (11) if it is an analytic function of the three spatial coordinates.

For models of galaxies with non-analytic cores we pick three from the one-parameter family studied by Dehnen (1993) and Tremaine et al. (1994). We call them "$\gamma$ models" because they are defined by the density which Dehnen writes as

$$\rho_\gamma(r) = \frac{(3-\gamma)}{4\pi} \frac{Ma}{r^\gamma(r+a)^{4-\gamma}}. \quad (12)$$

(Tremaine et al. describe the same density by the parameter $\eta = 3 - \gamma$ and call the models "$\eta$ models".) At large radii the density falls off as $\rho_\gamma(r) \sim r^{-4}$, just as for the isochrone model, but at small radii the density has a cusp $\rho_\gamma(r) \sim r^{-\gamma}$. The mass distribution $M_\gamma(r)$ is nonsingular as long as $\gamma < 3$:

$$M_\gamma(r) = M \left(\frac{r}{r+a}\right)^{3-\gamma}. \quad (13)$$

Two of these models are well known from previous work: the $\gamma = 1$ model is the Hernquist (1990) model;

---

[1] There is a typographical error in the Goodman-Binney paper suggesting that $\beta$ varies as $r$ and not $r^{1/2}$. The $x_m$ in their equation (13b) should be $x_m^2$. The same error is repeated by Binney and Petit (1989).

[2] Suggested to us by S. Tremaine.

$\gamma = 2$ is the Jaffe (1983) model. We use these as representative models for galaxies with mild ($\gamma = 1$) and steep ($\gamma = 2$) density cusps, and add as a third the $\gamma = 0$ model which, though it has a finite central density, does not qualify as a model with an analytic core because its density varies linearly with radius near the center. We have experimented with the $\gamma = 3/2$ model, but do not show the results here because they are intermediate between those for $\gamma = 1$ and $\gamma = 2$ and do not reveal any surprises.

The results from our calculations are shown in Figures 2–5. We present all the results (except those for the isothermal sphere in Fig. 1) in the standardized units of Heggie and Mathieu (1986), in which the gravitational constant $G$ and the initial galaxy mass $M$ are both chosen to be unity, and the initial energy is chosen to be $E = -1/4$. The scale lengths for the $\gamma$ and isochrone models are thus $a = 1/(5 - 2\gamma)$ and $b = (3\pi - 8)/6$. Each figure has four panels, showing (a) the surface density, (b) the projected velocity dispersion, (c) the anisotropy parameter $\beta$, and (d) the kurtosis of the LOSVD. Each panel of each figure has six lines: the dotted line shows the initial model before the BH is added; the five solid lines show the final models after the adiabatic growth of BHs of masses 0.001, 0.003, 0.01, 0.03, and 0.1.

### 3.2. Surface-density cusps

The surface-density cusps shown in Figures 2–5 vary from one model to another, and, for all models but the isochrone, are steeper than the $\Sigma \sim R^{-1/2}$ cusp found by Peebles (1972) and Young (1980) for the isothermal sphere. Varying the mass of the central BH merely shifts the radius where the limiting power-law cusp appears.

The fact that the $\gamma$ models develop cusps steeper than $\Sigma \sim R^{-1/2}$ is perhaps not surprising for the models with $\gamma > 0$, which have density cusps before the BH is added, but it is for the $\gamma = 0$ model, which starts with a finite central density as does the isochrone and yet develops a cusp that is twice as steep. There is a simple explanation for this difference. In deriving his cusp formula, Peebles (1972) assumed that the distribution function could be approximated by a constant in the core of the initial model. This assumption is not valid for the $\gamma$ models, because their distribution functions diverge as $E$ approaches $\phi(0)$. It is easy to generalize the derivation to take this into account (see Appendix A). We need just three assumptions: that the initial model has an isotropic core; that the potential varies with radius near $r = 0$ as a power law $\phi \sim r^{2-\gamma}$; and that the distribution function diverges near $E = \phi(0)$ as a power-law $f(E) \sim (E - \phi(0))^{-n}$. From these it follows that the adiabatic growth of a central BH induces a density cusp

$$\rho(r) \sim r^{-A}, \quad \Sigma(R) \sim R^{1-A},$$
$$A = \frac{3}{2} + n\left(\frac{2-\gamma}{4-\gamma}\right). \quad (14)$$

TABLE 1
Adiabatic Density Cusps

| Model | $\gamma$ | $n$ | $A$ | $C$ |
|---|---|---|---|---|
| isochrone | 0 | 0 | 3/2 | 9/4 |
| $\gamma = 0$ | 0 | 1 | 2 | 9/4 |
| $\gamma = 1$ | 1 | 5/2 | 7/3 | 7/3 |
| $\gamma = 3/2$ | 3/2 | 9/2 | 12/5 | 12/5 |
| $\gamma = 2$ | 2 | — | 5/2 | 5/2 |

For galaxy models with analytic cores, $n = 0$ and we recover the result $A = 3/2$, but for models with $n > 0$ we find steeper cusps. We have verified this prediction for the five models listed in Table 1. Equation (14) is not valid for $\gamma = 2$ (because the potential and distribution function do not behave as power laws near the center), but as $\gamma$ approaches 2 the cusp exponent $A$ for the $\gamma$ models approaches 5/2, which agrees with the numerical results shown in Figure 5.

There is a gap in Table 1 between the models with analytic cores, for which $A = 3/2$, and the $\gamma$ models, for which $A \geq 2$. There is another gap between the $\gamma$ models with $0 < \gamma < 2$, for which (Dehnen 1993)

$$n = \frac{6-\gamma}{2(2-\gamma)}, \quad (15)$$

and the model with $\gamma = 0$, for which $n = 1$. The gaps can be filled by other non-analytic models, such as the one-parameter family described by the density

$$\rho_\lambda(r) = \frac{\rho_0}{(r^\lambda + a)^{4/\lambda}}. \quad (16)$$

These models are awkward to work with because the potential and distribution function must be found by numerical integration. We experimented with some models with $1 < \lambda < 2$ and found final density cusps intermediate between $A = 3/2$ and $A = 2$. Equation (14) works in some cases but not all; it fails for the models with $\lambda$ close to (but less than) 2.

### 3.3. Velocity cusps

The cusps in velocity dispersion in Figures 2–5 all rise as $\langle v_p^2 \rangle \sim R^{-1}$ at small radii, and do not vary much from one model to another, although at low BH masses the cusp for the isochrone model is less noticeable than for the $\gamma$ models. Note that the velocity dispersion for the $\gamma = 1$ model without a BH (in fact, for any $\gamma$ model with $1 \leq \gamma < 2$) goes to zero at the center (see Binney 1980, Tremaine et al. 1994 for a discussion of this).

The anisotropy parameter $\beta$ behaves differently for the isochrone model than for the $\gamma$ models. The isochrone model remains isotropic at the center and develops a mild tangential anisotropy away from the center, similar to the result for the isothermal sphere shown in Figure 1. This is true also for the Plummer model and, we suspect, for all models with analytic cores. The $\gamma$ models develop larger tangential anisotropies that penetrate right to the center. Note also that increasing the mass of the central BH has a different effect for the two classes of models: for the isochrone model it increases the maximum anisotropy that develops; for the $\gamma$ models it does not change the maximum anisotropy, but merely shifts outward the radius to which the anisotropy reaches.

The results for the kurtosis of the LOSVD are more difficult to interpret. The figures show the deviation of the kurtosis from $\kappa = 3$, the expected value for a Gaussian distribution. In most cases this deviation is small in the final cusp, comparable in magnitude with what it was in the outer parts of the initial model without the BH. Note, however, the following differences between the behaviour of the kurtosis near the center of the isochrone model and the three $\gamma$ models: for the isochrone model without a BH the kurtosis is constant, whereas for the $\gamma = 1$ model (and other $\gamma$ models with $1 \leq \gamma < 2$) the kurtosis diverges; for the isochrone model the addition of a BH causes the kurtosis to increase, whereas for the three $\gamma$ models the opposite occurs.

Perhaps one conclusion to draw from the kurtosis plots is that the adiabatic growth of a central BH in a spherical galaxy does not cause the LOSVD to become highly non-Gaussian. But note that this conclusion applies to the LOSVD measured at one exact radius, i.e., to what could be observed if we had infinite resolution. The conclusion changes if the LOSVD is averaged over an aperture, because the average of $\langle v_p^4 \rangle$ will be weighted more towards the center ($R = 0$) than the average of $\langle v_p^2 \rangle$, and hence the effective kurtosis of the averaged LOSVD will differ from our comparison of $\langle v_p^4 \rangle$ and $\langle v_p^2 \rangle^2$ at the same radius. In fact, if the LOSVD is averaged over an aperture that includes the BH the effective kurtosis will be infinite, because of the arbitrarily high velocities possible close to the BH (see Bahcall and Wolf 1976). van der Marel (1994a) shows that this leads to a positive $h_4$ Gauss-Hermite moment, and stresses that it is better to quantify the observations by the Gauss-Hermite moments than by the classical moments such as $\langle v_p^2 \rangle$ and $\langle v_p^4 \rangle$. Our work shows at least that, for the models we consider, the non-Gaussian nature of the LOSVD observed near the center is expected to result almost entirely from the spatial averaging, and not from a peculiarity of the intrinsic velocity distribution.

## 4. Discussion

### 4.1. Theoretical questions

#### 4.1.1. Anisotropic initial conditions

The results presented above are for models that start with isotropic velocity distributions. We do not view this as a severe limitation. Young (1980) suggested that "in order to have an effect, the anisotropies must be significant inside the radius of influence $r_I \sim GM_H/\sigma_v^2$ of the black hole as reckoned in the unperturbed cluster." We have verified the correctness of this suggestion for Dejonghe's (1987) anisotropic Plummer models (we tried models with $q = \pm 1$, which have anisotropies $\beta(r) = 0.5qr^2/(1+r^2)$); even for the largest BH we considered ($M_{BH} = 0.1M$) the cusps were nearly identical with that for an isotropic Plummer model.

We also experimented with some models derived by Cuddeford's (1991) technique (with his $\alpha$ set to $1/2$) to have a constant tangential anisotropy $\beta(r) = -1/2$. For the isochrone and $\gamma = 0$ models the resulting cusps differed from those for the corresponding isotropic models, but not by much. That is not surprising, since we know the exact result for the adiabatic cusp that forms around a BH if the initial model has a density cusp $\rho(r) \sim r^{-\gamma}$ made up entirely of circular orbits (this is a simple generalization of Young's result, derived in Appendix A):

$$\rho(r) \sim r^{-C}, \quad \Sigma(R) \sim R^{1-C},$$
$$C = 3 - \frac{3-\gamma}{4-\gamma} \quad (17)$$

For the isochrone and $\gamma = 0$ models, this "circular" cusp slope is $C = 9/4$, considerably steeper than the "isotropic" cusp slope of $A = 3/2$ for the isochrone model, but not much steeper than the $A = 2$ for the $\gamma = 0$ model. For the $\gamma$ models with $0 < \gamma < 2$, the values of A and C coincide. Since this is for the most extreme tangential anisotropy possible, we conclude that a moderate tangential anisotropy in the initial model will have little or no effect on the final density cusp (although it will have an effect on the kinematics).

The case of a galaxy core with a radial anisotropy is more difficult to analyse because of a lack of suitable models to test. We tried to derive isochrone and $\gamma = 0$ models with a constant anisotropy $\beta = 1/2$ by Cuddeford's (1991) technique (with his $\alpha$ set to $-1/2$), but that was not possible: the distribution functions turned out to be negative at large binding energies. We do not know how large the radial anisotropy can be at the center of a galaxy with a flat core or a mild density cusp, but it appears to be small (O. Gerhard, private communication), probably too small to significantly change the cusp that forms around the central BH in our calculations.

4.1.2. *Origin of BH mass*

Young (1980) suggested that a central BH could grow "by accreting gas, from mass loss by giant stars, or by some other process," but he added a BH to his galaxy models without removing any mass from the surrounding stars. We have done some simple experiments to check how sensitive the results are to this assumption.

We first tried removing the BH mass uniformly from all the stars, by reducing the the stellar distribution function by a constant fraction $M_{BH}/M$ at the same time that we added the BH. This made almost no difference to the cusp around the BH, even for $M_{BH}/M$ as large as 0.1.

We then tried removing the BH mass from the stars of lowest angular momentum, since they are the stars that approach closest to the center. We adopted the following strategy: at the time the BH is added, reduce the distribution function $f(E, L)$ by the loss fraction $l_f$ if $L < L_l$, and leave it unchanged if $L \geq L_l$, i.e.,

$$f(E, L) \longrightarrow \begin{cases} (1 - l_f) f(E, L) & \text{if } L < L_l, \\ f(E, L) & \text{otherwise,} \end{cases} \quad (18)$$

with $L_l$ chosen so that the total mass removed equals $M_{BH}$. The rest of the calculation remains the same. We assume that, however the mass is lost, it is lost slowly so the action variables of the remaining stars are adiabatically conserved. The results of several calculations of this type for the $\gamma = 0$ model are shown in Figure 6.

The calculation with $l_f = 1.0$ yields a galaxy with a hidden BH, with no observable cusp in the surface density or projected velocity dispersion. In fact this galaxy model has a hole in the middle in two senses: a BH and a hole carved out of the intrinsic density $\rho(r)$. Although this is a contrived model — we have assumed that all of the mass with $L < L_l$ but none with $L > L_f$ gets swallowed by the BH, and have ignored perturbations such as two-body relaxation and triaxial components to the potential that might help replenish the density hole — it offers the intriguing possibility that real galaxies could contain BHs larger than suggested by their surface-density and velocity cusps. The calculations with $l_f < 1$ show, however, that the mass loss must be highly concentrated towards the center ($l_f \gtrsim 0.5$) for it to have a noticeable effect. The velocity-dispersion cusp is less sensitive to the effects of mass loss than is the surface-density cusp.

4.1.3. *Dynamical stability and related questions*

We are confident that our models are dynamically stable, but know of no theorem that proves this for non-isotropic models with a central BH. The question is especially interesting for models with hidden BHs, such as that in Figure 6 with $l_f = 1$. Another question is by how much our models would differ if the BHs grow too fast for the assumption of adiabatic invariance to be justified. We hope to answer these soon with the help of large N-body experiments.

4.2. **Observational implications**

Our results should be compared with observations with some caution. We have started our calculations from simple galaxy models that, while sharing the essential properties of real galaxies, are not expected to match them closely at all radii. We have assumed spherical symmetry and have ignored rotation, whereas many of the galaxies believed to contain massive BHs have rotating, disc-like nuclei. We have presented results that would be obtained with infinite resolution, and have ignored seeing corrections and other such factors that must be included in real-istic models. Despite these limitations, we believe our results lead to some important conclusions for the interpretation of density and velocity cusps in galactic nuclei.

*4.2.1. Models without BHs*

We start by asking how steep a cusp can be *without* a central BH. Dehen (1993) and Tremaine et al. (1994) describe $\gamma$ models without BHs for all $\gamma$ values between 0 and 3. Perhaps some of the models with large $\gamma$ values can be ruled out.

One restriction is set by the total energy, $E = -GM/4a(5-2\gamma)$, which diverges as $\gamma$ approaches $5/2$. Another is set by two-body relaxation. For any $\gamma$ model with $\gamma > 0$, the relaxation time goes to zero at the center. This sets a limit to the minimum radius for which the model can accurately represent a collisionless system (the radius at which $t_r$ equals the age of the system). For example, for the models with $1 < \gamma < 3$ we find, for $r \ll a$,

$$t_r(r) = \frac{0.065 v^3}{G^2 m \rho \ln \Lambda} \approx \frac{1.5}{(3-\gamma)(\gamma-1)^{3/2}} \left(\frac{a^3}{GM}\right)^{1/2} \left(\frac{M/m}{\ln \Lambda}\right) \left(\frac{M(r)}{M}\right)^{\frac{3-\gamma/2}{3-\gamma}}. \quad (19)$$

For $\gamma$ values close to 3 the exponent $(3-\gamma/2)/(3-\gamma)$ is large, and a substantial fraction of the mass is then at radii where $t_r$ is small enough to invalidate the assumptions on which the model is based (i.e., the mass would have undergone core collapse, possibly leading to the formation of a BH; see Quinlan and Shapiro 1990). While this argument does not give a sharp division between acceptable and unacceptable values for $\gamma$ (because it depends on the values of $M$, $m$, and $a$), $\gamma = 5/2$ seems a conservative choice.

If we therefore disregard models with $\gamma > 5/2$, we find that the steepest cusp possible without a BH is

$$\Sigma(R) \sim R^{-3/2}, \quad \langle v_p^2 \rangle \sim R^{-1/2} \quad (\gamma = 5/2). \quad (20)$$

This model has a steep surface-density cusp, much steeper than the adiabatic cusp around a BH in an isothermal sphere, but only a gradual velocity cusp, half as steep as expected around a BH. A surface-density cusp at the center of a galaxy thus provides only weak evidence for a BH; a Keplerian rise in the velocity dispersion towards the center is the real proof (if other sources of dark matter can be ruled out).

*4.2.2. Interpretation of density and velocity cusps*

Recent photometry of elliptical galaxies from the Hubble Space Telescope (HST) has shown almost none to have analytic cores (Crane et al. 1993; Ferrarese et al. 1994; Kormendy et al. 1994); most have surface brightnesses that continue to rise at the smallest radii resolved. The old picture of a galaxy having a core radius $r_c$ within which the surface brightness is nearly constant is giving way to a new picture where, for many galaxies, the inner parts can be modelled by a double power law such as

$$I(R) = 2^{(\alpha_2-\alpha_1)/\delta} I_b \left(\frac{r}{r_b}\right)^{-\alpha_1} \left[1 + \left(\frac{r}{r_b}\right)^\delta\right]^{(\alpha_1-\alpha_2)/\delta} \quad (21)$$

(from Kormendy et al. 1994, with the notation changed to avoid confusion with our use of $\gamma$), with the transition between the two powers $\alpha_1$ and $\alpha_2$ occuring at a radius $r_b$ called the "break" radius or "bend" radius or, sometimes (confusingly), the core radius. Paradoxically, the larger elliptical galaxies such as M87, the ones we expect to harbor massive BHs, have only gradual surface-density cusps ($\alpha_1 \simeq 0.0$–$0.3$), while the smaller ellipticals have the steepest cusps ($\alpha_1 \simeq 0.5$–$1.0$). The explanation for this dichotomy is not clear; it probably requires different formation mechanisms, perhaps involving massive BHs.

The double power law in equation (21) should not be confused with the density law (12) for the $\gamma$ models. These models (such as the Jaffe and Hernquist models) were designed to give reasonable fits to elliptical galaxies if the scale length $a$ is chosen comparable with the effective radius $r_e$. (Some of the BH masses in our calculations are therefore much larger than would be expected in real galaxies.) The break radius $r_b$ in equation (21) is typically a few arcseconds for Virgo-cluster ellipticals, at least 10 times smaller than $r_e$; and the power-law outside the break radius is typically $\alpha_2 \simeq 1.0$–$2.0$, much less steep than the $\sim R^{-3}$ fall off for a $\gamma$ model at $r > a$. The $\gamma$ models are not flexible enough to match closely a galaxy with a double power-law profile at $r \simeq r_b$.

Some galaxies have surface-brightness cusps that resemble Young's (1980) model for the adiabatic growth of a BH in an isothermal sphere. The best known of these is M87, first analysed in this way by Young et al. (1978). The HST photometry of M87 shows a gradual surface-density cusp, $I \sim R^{-0.26}$, which Lauer et al. (1992a) say is consistent with Young's model (with $M_{BH}$ as large as $3 \times 10^9 \, M_\odot$) because the observations do not penetrate to a small enough radius to see the expected asymptotic slope of $R^{-1/2}$. We checked this with our program and reached the same conclusion; observations with a slightly higher resolution should see a steepening in the surface brightness if Young's model is correct. M32 is another galaxy studied in detail (Lauer et al. 1992b) that can be fit by Young's model, although the fit is not as striking as it is for M87. Crane et al. (1993) fit Young-type cusps to a number of elliptical galaxies (and give convenient formulas for doing this), although they find that a single or double power-law often fits just as well, sometimes better (especially for galaxies with steep inner cusps).

These fits to Young's (1980) model are suggestive, but cannot be accepted as convincing evidence for massive BHs. In many cases it is easy to construct models without BHs that fit the data just as well. We caution against attaching too much significance to Young's $R^{-1/2}$ power-law in interpreting surface-brightness cusps; steeper cusps result from the adiabatic growth of BHs in galaxies with non-analytic cores, and the observations suggest that these are common. The adiabatic-growth scenario will always cause a steepening in the surface brightness at small radii (except in contrived models such as those in Fig. 6), although this is not so noticeable for a model like Jaffe's that starts with a steep density cusp.

Convincing evidence for massive BHs can come only from high-resolution spectroscopic observations. The velocity-dispersion cusp around a BH is insensitive to the details of the galaxy model within which the BH forms (unlike the surface-density cusp which varies from model to model), and is difficult or impossible to mimic without a BH. Ground-based observations of M87 show evidence for such a cusp (van der Marel 1994b), suggesting that the original Young et al. (1978) BH model is correct, a conclusion strengthened by the discovery of a high-velocity gas disk at the galaxy center (Harms et al. 1994). More results like these from the refurbished HST are eagerly awaited. When combined with surface photometry, they can help us assess models for the formation of massive BHs and dense galactic nuclei. There is still much work to do to refine these models so we can extract the most information possible from the observations.

We thank S. Faber, O. Gerhard, J. Kormendy, S. Tremaine, and R. van der Marel for helpful discussions while this work was in progress. This work was supported in part by the Alfred P. Sloan Foundation, NASA Theory Grant NAGW–2422, and from the NSF under Grants AST 90–18526, ASC 93-18185, and the Presidential Faculty Fellows Program.

## A. Adiabatic density cusp around a BH

In the derivation that follows we drop all inessential variables and numerical constants ($G$, $\pi$, $M$, etc.) and consider only the scaling of various quantities with $E$ and $r$. We use subscripts $i$ and $f$ where appropriate to distinguish the initial and final states, and adopt a sign convention where $E$ and $\phi$ are both positive. We approximate the final potential by the Kepler potential around the BH.

Consider first a model that starts with an isotropic core. The distribution function $f(E)$ is related to the differential energy distribution $N(E)dE$ (the number of stars with energies in the range $E$ to $E + dE$) by

$$N(E) = g(E)f(E), \qquad (A1)$$

where the density of states $g(E)$ is, for the isotropic initial model,

$$g_i(E_i) \sim \int_0^{\phi_i^{-1}(E_i)} dr \, r^2 \sqrt{\phi_i(r) - E_i} \sim E_i^{(8-\gamma)/2(2-\gamma)}. \qquad (A2)$$

Near the BH in the final model the energy varies with radius as $E_f \sim 1/r$, which allows us to relate the density $\rho_f(r)$ to $N_f(E_f)$ and $N_i(E_i)$ by

$$\rho_f(r) \sim r^{-2} N_f(E_f) \left(\frac{dE_f}{dr}\right) \sim r^{-4} N_i(E_i) \left(\frac{dE_i}{dE_f}\right). \qquad (A3)$$

The relation between $E_i$ and $E_f$ is easy to derive for purely circular orbits and purely radial orbits, for which the invariance of the action (angular momentum in the circular case, radial action in the radial case) implies that

$$E_i^{(4-\gamma)/2(2-\gamma)} \sim E_f^{-1/2}, \qquad (A4)$$

or

$$E_i \sim E_f^{-(2-\gamma)/(4-\gamma)} \sim r^{(2-\gamma)/(4-\gamma)}. \qquad (A5)$$

If we use this relation for all orbits we find from equation (A3) that

$$\rho_f(r) \sim r^{-A}, \quad A = \frac{3}{2} + n\left(\frac{2-\gamma}{4-\gamma}\right). \qquad (A6)$$

A model consisting entirely of circular orbits is easier to consider. Assume that the density cusp is $\rho \sim r^{-\gamma}$ before the addition of the BH and $\rho \sim r^{-C}$ after. Conservation of mass implies that

$$\rho_i r_i^2 \, dr_i = \rho_f r_f^2 \, dr_f \quad \Longrightarrow \quad r_i^{3-\gamma} \sim r_f^{3-C}. \quad (A7)$$

Conservation of angular momentum implies that

$$r_i M_i(r) = r_f M_f(r) \simeq r_f M_{BH} \quad \Longrightarrow \quad r_i^{4-\gamma} \sim r_f. \quad (A8)$$

Combining these two results we find

$$C = 3 - \frac{3-\gamma}{4-\gamma}. \quad (A9)$$

**Figure Captions**

Fig. 1.— Anisotropy induced by the growth of a central black hole in an isothermal sphere, as quantified by (a) the ratio $f(E, Lc(E))/f(E, 0)$, and (b) the anisotropy $\beta$ in the velocity dispersion. The results are from a numerical calculation with a self-consistent potential (solid curve), from the same calculation with an idealized, non-self-consistent potential (dashed curve), and from the approximate solution of Goodman and Binney (filled squares). See text for details.

Fig. 2.— Results from the adiabatic growth of central black holes of masses 0.001, 0.003, 0.01, 0.03, and 0.1 in an isochrone model (the dotted lines show the initial model without a black hole): (a) surface density; (b) line-of-sight velocity dispersion; (c) anisotropy in the velocity dispersion; (d) kurtosis (minus three) of the line-of-sight velocity distribution.

Fig. 3.— Results from the adiabatic growth of central black holes in a $\gamma = 0$ model (see Fig. 2).

Fig. 4.— Results from the adiabatic growth of central black holes in a $\gamma = 1$ (Hernquist) model (see Fig. 2).

Fig. 5.— Results from the adiabatic growth of central black holes in a $\gamma = 2$ (Jaffe) model (see Fig. 2).

Fig. 6.— Adiabatic cusp induced by the growth of a central black hole of mass 0.01 in a $\gamma = 0$ model (cf. Fig. 3) when the black hole grows at the expense of low angular-momentum stars. The fractional mass loss is $l_f = 1.0$ (bottom curves), 0.9, 0.5, and 0.0 (no mass loss; top curves). The dotted curves show the initial model without a black hole.